\documentclass[11pt, letterpaper, oneside]{article}

\usepackage{graphicx}
\usepackage[body={6.6in, 9in}, right=1in, top=1in]{geometry}
\usepackage{amssymb}
\usepackage{amsmath} 
\usepackage{feynmp} 
\usepackage{subfigure}
\usepackage{slashed}
\usepackage{array}
\usepackage{bm}
\usepackage[linktoc=page,colorlinks=true,linkcolor=blue,citecolor=blue,urlcolor=blue]{hyperref}

\newcommand{\be}{\begin{equation}}
\newcommand{\ee}{\end{equation}}
\newcommand{\fields}{\underline{\phi}}

\newcolumntype{M}[1]{>{\centering\arraybackslash}m{#1}}

\def\be{\begin{equation}}
\def\ee{\end{equation}}
\def\ba{\begin{eqnarray}}
\def\ea{\end{eqnarray}}

\newcommand{\sfrac}[2]{{\textstyle\frac{#1}{#2}}}

\begin{document}

\begin{fmffile}{graphs}

% TITLE AND AUTHOR

\begin{center}
\Large{\textbf{An improved Noether's theorem for spacetime symmetries}} \\[1cm]
\large{Ioanna Kourkoulou, Alberto Nicolis, and Guanhao Sun}
\\[0.4cm]

\vspace{.2cm}
\small{\em Center for Theoretical Physics and Department of Physics, \\
  Columbia University, New York, NY 10027, USA}

\end{center}

\vspace{.2cm}

% ABSTRACT

\begin{abstract}
We exploit an ambiguity somewhat hidden in Noether's theorem to derive systematically, for relativistic field theories, the stress-energy tensor's improvement terms that are associated with additional spacetime symmetries beyond translations. We work out explicitly the cases of Lorentz invariance, scale invariance, and full conformal invariance.
The main idea is to use, directly in the translation Noether theorem, the fact that these additional symmetries can be thought of as suitably modulated translations. Compared to more standard derivations of the improvement terms, ours (1) unifies all different cases in a single framework, (2) involves no guesswork, (3) yields the desired algebraic properties (symmetry and/or tracelessness) of the stress-energy tensor {\em off-shell}, and (4) unifies the translation Noether theorem with those of the additional spacetime symmetries, yielding at the same time both the improved stress-energy tensor and the additional Noether currents.

\end{abstract}

%\newpage

% TABLE OF CONTENTS

%\tableofcontents

%\newpage

% INTRODUCTION

%%%%%%%%%%%%%%%%%%%%%%%%%%%%%%%%%%%%%%%%%%%
%%%%%%%%%%%%%%%%%%%%%%%%%%%%%%%%%%%%%%%%%%%

\section{Introduction: Ambiguous currents from an ambiguous theorem}

Noether's theorem relates continuous symmetries to conservation laws. For local field theories, it yields one locally conserved current for each independent symmetry generator. However, the conserved currents one derives from the theorem are notoriously ambiguous, for two main reasons: 
\begin{enumerate}
\item
First, given a current $J^\mu$ that is conserved ``on-shell"---that is, on solutions of the equations of motion---one can always add to it a contribution of the form
\be \label{dSigma}
\Delta J^\mu \equiv \partial_\alpha \Sigma^{\alpha \mu} \; , \qquad \qquad 
\Sigma^{\alpha\mu} = - \Sigma^{\mu\alpha} \; ,
\ee 
where $\Sigma^{\alpha \mu}$ is any local functional of the fields that is antisymmetric in $\alpha$ and $\mu$. Such an addition is conserved ``off-shell"---that is, on any field configuration, regardless of whether this solves the equations of motion or not. Moreover, it does not contribute to the global charge associated with the current, $Q \equiv \int d^3 x J^0$, because, using the antisymmetry of $\Sigma^{\alpha \mu}$ and assuming the fields vanish sufficiently fast at spatial infinity, one has
\be
\int d^3 x \, \Delta J^{0} = \int d^3 x \, \partial_i \Sigma^{i 0} = 0 \; .
\ee
So, the two currents, $J^\mu$ and $J'^\mu = J^\mu + \Delta J^\mu$, obey equivalent conservation laws and yield the same global charge. 
\item 
Second, since the current $J^\mu$ is only conserved  on-shell anyway, one can add to it contributions that vanish on-shell. At the classical level, this does not modify the value of $J^\mu$ or of $Q$ on solutions of the equations of motion, and so it does not modify the associated conservation laws either. At the quantum level, this  modifies the Ward identities in the contact terms only, since the equations of motion are obeyed in correlation functions  up to contact terms.
\end{enumerate}
These two ambiguities are routinely exploited in the case of spacetime symmetries, to ``improve" the current associated with spacetime translations, the stress-energy tensor $T^{\mu\nu}$: if the theory has spatial rotational invariance, $T^{ij}$ can be made symmetric; if the theory also has Lorentz invariance, the full $T^{\mu\nu}$ can be made symmetric \cite{Belinfante, Weinberg1}; if the theory further has scale-invariance, $T^{\mu\nu}$ can be made traceless up to a total divergence \cite{Wess, CCJ, CJ}; finally, if the theory has full conformal symmetry, $T^{\mu\nu}$ can be made fully traceless \cite{Wess, CCJ, CJ}. Notice that these algebraic properties of symmetry and tracelessness are generically only valid on-shell; however, we can always make them valid off-shell by adding suitable terms of the type discussed in item 2 above.

For the purposes of what follows, it is instructive to trace the ambiguities discussed above back to Noether's theorem: it is the theorem itself that is ambiguous. To see this, let's review how the theorem usually works. From now on, we will be using $\fields(x)$ to denote a generic multiplet of (real) fields, not necessarily scalars, and the dot `$\,\cdot\,$' to denote the contraction of indices in field space.
We assume that we have an action $S[ \phi ]$ that is invariant under some continuous symmetries, enumerated by $a$,
\be
\fields \to \fields  + \epsilon^a \, \underline{\Delta}_a  \; , \label{sym}
\ee
where $\epsilon^a$ are constant infinitesimal parameters, which we will always keep up to first order only, and $\underline{\Delta}_a$ are given local functionals of the fields. To be precise, let's say that under \eqref{sym} the Lagrangian density changes by a total derivative,
\be \label{TD}
{\cal L} \to {\cal L} + \epsilon^a \, \partial_\mu F^\mu_a \; ,
\ee 
where $F^\mu_a$ are some functionals of the fields.
Then, the theorem goes, let's see how the Lagrangian density changes if we make $\epsilon^a$ in \eqref{sym} spacetime-dependent, $\epsilon^a = \epsilon^a(x)$. To first order in $\epsilon^a$, the variation must take the form
\be
\delta {\cal L} = \epsilon^a (x) \, \partial_\mu F^\mu_a + \partial _\mu \epsilon^a (x) \, G^\mu_a + \partial _\mu \partial_\nu \epsilon^a (x) \, G^{\mu\nu}_a + \dots \,
\ee
where the $G$'s are suitable functionals of the fields, and usually the series in derivatives of $\epsilon^a$ truncates at finite 
order. For example, for a Lagrangian with at most $N$ derivatives on a single field, the series usually truncates at $\partial^N \epsilon^a$ order. Notice that if one sets $\epsilon^a$ to a constant, only the first term survives, and one goes back to eq.~\eqref{TD}. If we now integrate $\delta {\cal L}$ over spacetime we get the variation of the action for generic $\epsilon^a(x)$. Restricting to functions $\epsilon^a(x)$ that go to zero at infinity, we can integrate all  derivatives of $\epsilon^a$ by parts  and end up with
\be \label{delta S}
\delta S = \int d^4 x \,\delta {\cal L} =  \int d^4 x \, \epsilon^a(x) \, \partial_\mu J^\mu_a \qquad \qquad \mbox{(off-shell)} \; ,
\ee
%where $J^\mu$ is a suitable functional of the fields. 
where $J^\mu_a$ are whatever functionals of the fields emerge from the procedure just described. This defines the Noether currents. The last step is to recognize that eq.~\eqref{sym} for generic $\epsilon^a(x)$ vanishing at infinity is a particular field variation that vanishes at infinity, but on-shell the action should be stationary for {\em all} field variations that vanish at infinity. So, on-shell one must have
\be
\partial_\mu J^\mu_a = 0 \qquad \qquad \qquad \qquad \mbox{(on-shell)}\; .
\ee

Why are we saying that such a procedure is ambiguous? The ambiguity of item 1 above is easy to spot: Adding to $J^\mu_a$ identically conserved terms of the form \eqref{dSigma} does nothing to the integrand in \eqref{delta S}, precisely because such terms are identically conserved. The ambiguity of item 2 instead is more subtle to unveil, but more relevant for what follows. It has to do with the very first step of the Noether procedure, when we make $\epsilon^a$ spacetime dependent: it is usually assumed that that corresponds to replacing \eqref{sym} simply with
\be
\fields \to \fields  + \epsilon^a(x) \, \underline{\Delta}_a  \; , \label{sym2}
\ee
but, in fact, in the theorem as we just described it, nowhere are we using this specific form of the transformation. The only property that we are using is that the $x$-dependent transformation that we perform should reduce to the symmetry \eqref{sym} in the limit in which $\epsilon^a$ are constants. Then, instead of \eqref{sym2}, we could use \cite{KNS}
\be
\fields \to \fields  + \epsilon^a(x) \, \underline{\Delta}_a   + \partial_\mu \epsilon^a(x) \, \underline{\Phi}^\mu_a +
 \partial_\mu \partial_\nu  \epsilon^a(x) \, \underline{\Phi}^{\mu\nu}_a + \dots \label{sym3} \; ,
\ee
where the $\underline{\Phi}$'s are {\em arbitrary} functionals of the fields. By definition of functional derivatives, these new terms in the transformation of $\fields$ modify the variation of the action \eqref{delta S} by
\be
\delta S \supset  \int d^4 x \, \frac{\delta S}{\delta \fields} \cdot \big[ \partial_\mu \epsilon^a(x) \, \underline{\Phi}^\mu_a +
 \partial_\mu \partial_\nu  \epsilon^a(x) \, \underline{\Phi}^{\mu\nu}_a + \dots \big] \qquad \qquad \mbox{(off-shell)} \; .
\ee
Integrating by parts all derivatives of $\epsilon^a$ and comparing to \eqref{delta S}, we see that the current gets new contributions of the form
\be
J^\mu_a \supset - \frac{\delta S}{\delta \fields} \cdot \big[  \underline{\Phi}^\mu_a - \partial_\nu  \underline{\Phi}^{\mu\nu}_a + \dots \big] \qquad \qquad \mbox{(off-shell)} \; ,
\ee
which clearly vanish on-shell.

So, in summary, both ambiguities discussed above are inherent features of  Noether's theorem itself. The second one---the one we just discussed---is more interesting, in that it ties new terms in the current to a modification of how the fields are declared to transform under the spacetime-modulated version of the symmetry. This will allow us to derive systematically, directly from the theorem, the improvement terms for the stress-energy tensor that are associated with spacetime symmetries beyond translations.
As we now explain, the main idea is to tailor, each time, the {\em translation} Noether theorem to the particular {\em additional} symmetry one wants to exploit.

\vspace{.5cm}
\noindent
{\em Note added in revised version}:
Brauner, Torrieri, and Yonekura have brought to our attention refs.~\cite{Brauner:2014aha, Brauner:2019lcb, Yonekura:2012uk}, which also explore ambiguities in Noether's theorem and use them to derive, in a systematic way, a number of desired properties for the conserved currents. The overlap of our work with those papers is, in fact,  substantial, especially with ref.~\cite{Brauner:2019lcb}. We plan to analyze the connection more closely in the near future.

\vspace{.5cm}
\noindent
{\em Notation and conventions}:
We use natural units ($\hbar = c=1$) and the `mostly minus' signature for the metric throughout. 
For Lorentz generators, we use the same normalization as \cite{CCJ, CJ}, which differs from that of Weinberg \cite{Weinberg1}, $\mathcal{J}^{\mu\nu}_{\rm here} = - i \, \mathcal{J}^{\mu\nu}_{\rm Weinberg}$. When addressing scale- and conformal-invariance, we keep the spacetime dimensionality $D$ generic. Otherwise, we work in $D=4$.

%%%%%%%%%%%%%%%%%%%%%%%%%%%%%%%%%%%%%%%%%%%%%%%
%%%%%%%%%%%%%%%%%%%%%%%%%%%%%%%%%%%%%%%%%%%%%%%
\section{The main idea}

We want to improve the Noether's theorem for spacetime translations, exploiting the ambiguities discussed above, especially the second one. This, as we saw, is related to modifying the transformation properties of the fields as in \eqref{sym3}. We then have to ask under what condition we can  gain something by considering \eqref{sym3}, perhaps with some judicious choice of the $\underline{\Phi}$ functionals, instead of the apparently simpler \eqref{sym2}. 

In fact, for internal symmetries, we see no general benefit of using \eqref{sym3} in place of \eqref{sym2}. 
On the other hand, for translations, we can make use of the fact that other possible spacetime symmetries, such as Lorentz invariance, scale invariance, and conformal invariance, can in fact be thought of as very specific spacetime-modulated translations. 
For instance, an infinitesimal Lorentz transformation of constant parameter $\omega_{\mu\nu} = -\omega_{\nu\mu}$, shifts the coordinates by
\be
x^\mu \to x'^\mu = x^\mu + \omega^\mu{}_\nu x^\nu \qquad\qquad \mbox{(Lorentz)}  \; ,
\ee
which is formally a spacetime modulated translation
\be
x^\mu \to x'^\mu = x^\mu + \epsilon^\mu(x) \; ,
\ee
with parameter
\be \label{Lorentz epsilon}
\epsilon^\mu(x) = \epsilon_L^\mu(x) \equiv \omega^\mu{}_\nu \, x^\nu
\ee
(`$L$' for `Lorentz'.) Notice that the index $a$ of the infinitesimal parameters now becomes the spacetime index, $\epsilon^a \rightarrow \epsilon^{\mu}$.

Our fields transform under a translation of constant parameter $\epsilon^\mu$ as, 
\be \label{translate}
\underline{\phi}(x) \; \rightarrow \; \underline{\phi}(x)- \epsilon^{\mu}\partial_{\mu}\underline{\phi}(x) \qquad\qquad \mbox{(translations)} \; ,
\ee
and usually we would run the translation Noether theorem by generalizing this to
\be \label{standard}
\underline{\phi}(x) \; \rightarrow \; \underline{\phi}(x)- \epsilon^{\mu}(x) \, \partial_{\mu}\underline{\phi}(x) \; ,
\ee
for generic $\epsilon^\mu(x)$, which yields the so-called canonical stress-energy tensor. But, in alternative, we can notice
that since under a Lorentz transformation of constant parameter $\omega_{\mu\nu}$ the fields transform as
\be \label{Lorentz dphi}
\underline{\phi}(x) \; \rightarrow \; 
\underline{\phi}(x) - \omega^{\mu} {}_\nu \, x^\nu \partial_{\mu} \underline{\phi}(x) - \sfrac{1}{2}\omega_{\mu\nu} \,\mathcal{J}^{\mu\nu} \cdot \underline{\phi}(x) \qquad\qquad \mbox{(Lorentz)} 
\; ,  
\ee
we can run the Noether's theorem for translations with a ``Lorentz-friendly" version of \eqref{sym3}:
\be \label{LI}
\underline{\phi}(x) \; \rightarrow \; \underline{\phi}(x) - \epsilon^{\mu}(x) \partial_{\mu} \underline{\phi}(x) - \sfrac{1}{2}\partial_{\mu}\epsilon_{\nu}(x) \mathcal{J}^{\mu\nu} \cdot \underline{\phi}(x) \; .  
\ee
Such a transformation rule has the property that for a very specific $\epsilon^\mu(x)$---eq.~\eqref{Lorentz epsilon}---it corresponds to a symmetry of the action: Lorentz invariance.
This implies that the stress-energy tensor one derives running the Noether's theorem in this way will have an additional property besides conservation. In particular, as we will see, the fact that eq.~\eqref{Lorentz epsilon} corresponds to the most general $\epsilon^\mu(x)$ that has constant, antisymmetric first derivatives, will force  the stress-energy tensor  to be automatically {\em symmetric}, off-shell. 

This of course matches the standard conclusion---Lorentz implies a symmetric $T^{\mu\nu}$---but in our procedure there is no guesswork involved: by making the translation Noether theorem sensitive to Lorentz invariance, we automatically get a symmetric stress-energy tensor. And, contrary to the standard Belinfante procedure, nowhere do we have to use the equations of motion. Of course, this means that the Belinfante stress-energy tensor differs from ours by terms proportional to the equations of motion. As we tried to emphasize in the Introduction, such an ambiguity is to be expected on general grounds.

This is the basic idea that we will try to exploit. As usual, the devil is in the details, so let's see explicitly how these work out in the case of Lorentz, scale, and conformal invariance.   

%%%%%%%%%%%%%%%%%%%%%%%%%%%%%%%%%%%%%%%%%%%%%%%
%%%%%%%%%%%%%%%%%%%%%%%%%%%%%%%%%%%%%%%%%%%%%%%
\section{Passive vs.~active, action vs.~Lagrangian}\label{passive vs active}

As a preliminary step, it is useful to be precise about the symmetries we want to consider. At least for the cases we study here, we can think of a spacetime symmetry as a symmetry of the action that is associated with a specific change of coordinates, 
\be
x^\mu \to x'^\mu(x)
\ee
under which the fields transform in a certain way,
\be
\fields (x) \to \fields'(x') \; .
\ee
This is the so-called passive viewpoint. According to it, the transformation above is a symmetry if the infinitesimal action element does not change
\be
 {\cal L}[\phi'(x')] \, d^4 x' = {\cal L}[\phi(x)] \, d^4 x\; ,
\ee
that is, if the Lagrangian density changes as
\be \label{action symmetry}
{\cal L}[\phi'(x')] = \Big| \det \frac{\partial x}{\partial x'} \Big|  {\cal L}[\phi(x)]
\ee
In principle we could insist on a weaker requirement---that the action change only by boundary terms---but, at least for standard spacetime symmetries, this subtlety happens to be relevant only in the case of conformal transformations and we will discuss it in due time.  For the moment, we are going to ignore it.  

Now, it so happens that, for an infinitesimal transformation, it is more convenient to adopt the so-called active viewpoint, whereby we transform directly the fields, evaluating all fields and derivatives at the same values of their arguments. Writing
\be
x'^\mu(x) = x^\mu + \epsilon^\mu(x) \; , 
\ee
for the specific $\epsilon^\mu(x)$ that corresponds to the infinitesimal symmetry we want to consider, from 
\eqref{action symmetry} we get
\be
{\cal L}[\phi'(x)] + \epsilon^\mu \partial_\mu {\cal L} = {\cal L}[\phi(x)] - \partial_\mu \epsilon^\mu {\cal L} \; ,
\ee
where we kept terms up to first order in $\epsilon^\mu$.
Thus, from the active viewpoint, the spacetime transformation under study is a symmetry of the action if the Lagrangian density changes by a {\em specific} total derivative term \cite{galilei}:
\be \label{d xi L}
\delta{\cal L} \equiv {\cal L}[\phi'(x)] - {\cal L}[\phi(x)]  = - \partial_\mu\big (\epsilon^\mu \,{\cal L} \big) \; .
\ee
We are now ready to run our improved translation Noether theorems.

\section{Improved translation Noether theorem}

\subsection{Generalities}\label{generalities}
For simplicity, we  focus our attention on Poincar\'e invariant field theories, with the field multiplet $\fields$ transforming
linearly under Lorentz symmetry, according to a generic representation $\mathcal{J}^{\mu\nu}$, not necessarily irreducible. Moreover, when we consider scale-invariant or conformal-invariant theories, we  assume that the fields transform linearly under those as well. In other words, we assume that no spacetime symmetry is spontaneously broken. As we hope it will be clear shortly, these simplifying assumptions are not really needed for our strategy to work, and in principle, our analysis can be straightforwardly extended to non-linear realizations as well.

Likewise, for simplicity we consider Lagrangian densities that depend at most on the first derivatives of the fields,
\be
{\cal L}[\phi] = {\cal L}(\phi ,\partial_\mu \phi) \; . 
\ee
In principle we could repeat our analysis with any number of higher derivatives. Even better, in this day and age, given the ongoing proliferation of effective field theories and derivative expansions, it would be nicer to find a more general functional approach that does not require specifying the maximum number of derivatives entering the Lagrangian. We leave this task for future work.

Under rigid translations, our fields transform as in \eqref{translate}. We want to run the associated Noether's theorem generalizing that transformation law to
\be \label{general}
\underline{\phi}(x) \rightarrow \underline{\phi}(x) - \epsilon^{\mu}(x) \partial_{\mu} \underline{\phi}(x) - \partial_{\mu}\epsilon_{\nu}(x) \underline{\Psi}^{\mu\nu} (x) \; ,  
\ee
for generic $\epsilon^\mu(x)$, and for a specific (field-dependent) $\underline{\Psi}^{\mu\nu} (x)$, which will change from case to case, depending on the additional symmetries we want to consider.

The variation of the Lagrangian density is
\be
\delta {\cal L} = \frac{\partial {\cal L}}{\partial \fields} \cdot \delta \fields + \frac{\partial {\cal L}}{\partial \,\partial_\alpha \fields} \cdot \partial_\alpha \delta \fields  \; ,
\ee
which, after some straightforward algebra, can be written as
\be
\label{general dL} \delta \mathcal{L}  = -\partial_{\mu}\left( \epsilon^{\mu}\mathcal{L} \right) - \partial_{\mu}\epsilon_{\nu} \: {\cal T}^{\mu\nu}- \partial_{\rho}\partial_{\mu}\epsilon_{\nu}  \: {\cal S}^{\rho\mu\nu} 
\ee
with
\begin{align}
\label{taumn} {\cal T}^{\mu\nu} & = T^{\mu\nu}_c +  \frac{\delta S}{\delta \underline{\phi}} \cdot \underline{\Psi}^{\mu\nu} +  \partial_{\rho} \, {\cal S}^{\rho\mu\nu} \\
\label{Srmn}{\cal S}^{\rho\mu\nu} & =  \frac{\partial\mathcal{L} }{ \partial \, \partial_{\rho}\underline{\phi}  } \cdot \underline{\Psi}^{\mu\nu} \; ,
\end{align}
where $T^{\mu\nu}_c$ is the so-called canonical energy-momentum tensor---the one that emerges from the standard Noether procedure---
\be
T_c^\mu {}_\nu = \frac{\partial\mathcal{L} }{ \partial  \, \partial_{\mu} \underline{\phi}} \cdot \partial_\nu \fields - \delta^\mu_\nu {\cal L} \; ,
\ee
and the equations of motion $\frac{\delta S}{\delta \underline{\phi}}$ are nothing but the Euler-Lagrange equations,
\be
\frac{\delta S}{\delta \underline{\phi}} =   \frac{\partial\mathcal{L} }{ \partial \underline{\phi}}  -\partial_\mu \bigg(\frac{\partial\mathcal{L} }{ \partial  \, \partial_{\mu} \underline{\phi}} \bigg) \; .
\ee
Also notice that the quantity ${\cal T}^{\mu\nu}$ differs from the canonical stress-energy tensor, $T^{\mu\nu}_c$, by two terms: one is proportional to the equations of motion, the other is a total derivative. As a result, at low-energies ${\cal T}^{\mu\nu}$ reduces to $T^{\mu\nu}_c$ on-shell. 

According to the general logic of Noether's theorem as reviewed in the Introduction, for {\em generic} $\epsilon^\mu(x)$ the variation of the Lagrangian density must take the form
\be \label{define Tmn}
\delta {\cal L} = - \partial_\mu \epsilon_\nu \, T^{\mu\nu} + \mbox{total derivatives} \; ,
\ee
and this can be taken as the definition of the stress-energy tensor. To rewrite \eqref{general dL} as \eqref{define Tmn}, we must integrate by parts the third term. 
The obvious way to do this would be to write
\be
- \partial_{\rho}\partial_{\mu}\epsilon_{\nu} \: {\cal S}^{\rho\mu\nu} = 
\partial_{\mu}\epsilon_{\nu}  \: \partial_{\rho}{\cal S}^{\rho\mu\nu}
+ \mbox{total derivatives} \; ,
\ee
but, in fact, we can be more general and keep in mind an ambiguity related to that of item 1 in the Introduction: since $\partial_\rho \partial_\mu \epsilon_\nu$ is symmetric in $\rho$ and $\mu$, we can add to whatever multiplies it any functional of the fields that is antisymmetric in $\rho$ and $\mu$.
This will prove useful for what follows. We can thus write that, for our generalization of the translation Noether theorem, the stress-energy tensor is
\be \label{general Tmn}
T^{\mu\nu} = {\cal T}^{\mu\nu} - \partial_{\rho}\Big(  {\cal S}^{\rho\mu\nu} + \Sigma^{\rho\mu\nu} \Big)  \; , \qquad \qquad 
\Sigma^{\rho\mu\nu} = - \Sigma^{\mu\rho\nu} \; ,
\ee
where $\Sigma^{\rho\mu\nu}$ is a generic functional of the fields that is antisymmetric in $\rho$ and $\mu$.

Notice that, crucially, to arrive to \eqref{general dL} we did {\em not} drop total-derivative terms. So, we can use the result of the last section: for the choices of $\epsilon^\mu(x)$ and $\underline{\Psi}^{\mu\nu}(x)$ that make \eqref{general} a {\em symmetry} transformation, only the first term in \eqref{general dL} should survive. As we'll see, this will typically imply some algebraic property for ${\cal T}^{\mu\nu}$. One can then try to use the ambiguity associated with $\Sigma^{\rho\mu\nu}$ to extend that property to the full $T^{\mu\nu}$
\footnote{For this reason, one should resist the temptation to cancel the last term in ${\cal T}^{\mu\nu}$  (see eq.~\eqref{taumn}) against the first term inside the $\partial_\rho(\dots)$ in eq.~\eqref{general Tmn} until all the symmetries have been exploited.}.
It may seem that at this point our procedure  could use some guesswork, despite our bragging about the opposite.
In practice, however, one parametrizes $\Sigma^{\rho\mu\nu}$ as the most general linear combination of the tensors at one's disposal
%---${\cal S}$, its transposed versions, and its trace parts in the general case, plus one other tensor, its transposed versions, and its trace parts in the conformal case---
with the right symmetries, and checks whether there is a choice of coefficients that achieves the desired result. Phrased in this way, this step is a linear algebra problem and, as advertised, there is no guesswork involved. We will see this explicitly at work in the three examples that follow.

%%%%%%%%%%%%%%%%%%%%%%%%%%%%%%%%%%%%%%%%%%%
%%%%%%%%%%%%%%%%%%%%%%%%%%%%%%%%%%%%%%%%%%%
\subsection{Lorentz-friendly version}\label{Lorentz friendly}
%The existing derivations of the Belinfante tensor rely mostly on an ad-hoc procedure of adding the said improvement terms that work nicely for symmetrizing the tensor on-shell while preserving its properties. In this paper we attempt to construct a more systematic approach for the emergence of these terms and take things a step further. Namely, we will show in this section a procedure that produces stress energy tensors that are symmetric even off-shell. 
%
To begin with, consider a Poincar\'e-invariant field theory, with the fields $\fields$ transforming according to some representation $\mathcal{J}^{\mu\nu}$ under Lorentz, as in eq.~\eqref{Lorentz dphi}. So, if we choose the functionals $\underline{\Psi}^{\mu\nu}$ in \eqref{general} to be simply
\be \label{Psi Lorentz}
\underline{\Psi}^{\mu\nu}(x) = \underline{\Psi}_L^{\mu\nu}(x) \equiv  \sfrac12 \mathcal{J}^{\mu\nu} \cdot \underline{\phi}(x)
\ee
(`$L$' for `Lorentz'), 
we have that for $\epsilon^\mu(x) = \epsilon_L(x) \equiv \omega^\mu {}_\nu x^\nu$, with constant and antisymmetric $\omega_{\mu\nu}$, eq.~\eqref{general} corresponds to a symmetry transformation, that is, only the first term in \eqref{general dL} should survive.

On the other hand, for these specific choices, eq.~\eqref{general dL} reads
\be \label{LorTran}
\delta_L {\cal L} = - \partial_\mu (\epsilon_L^\mu {\cal L}) - \omega_{\mu\nu} {\cal T}^{\mu\nu} \; .
\ee
This immediately implies that ${\cal T}^{\mu\nu}$ is symmetric, since $\omega_{\mu\nu}$ is the most general antisymmetric constant tensor:
\be
{\cal T}^{\mu\nu} = {\cal T}^{\nu\mu} \; .
\ee
Keep in mind that ${\cal T}^{\mu\nu}$ is not conserved by itself, and has to be complemented by a correction $\Delta T^{\mu\nu}$ in order to restore conservation. The reason behind this is that \eqref{LorTran} is obtained using a specific transformation, while in \eqref{define Tmn} we are using an arbitrary $\epsilon^\mu$. We should then ask whether we can choose $\Sigma$ in \eqref{general Tmn} to make the rest of the stress-energy tensor,
\be \label{rest}
\Delta T^{\mu\nu} = - \partial_{\rho}\Big( {\cal S}^{\rho\mu\nu} + \Sigma^{\rho\mu\nu} \Big) \; ,
\ee
also symmetric. We parametrize $\Sigma$ as the most general linear-in-${\cal S}$ combination of ${\cal S}$ and $\eta$ tensors with the correct anti-symmetry property. Taking into account that ${\cal S}^{\rho\mu\nu}$ is, in this case, antisymmetric in $\mu\nu$ (because of \eqref{Srmn} and \eqref{Psi Lorentz}), we write
\be
\Sigma^{\rho\mu\nu} = \Sigma_L^{\rho\mu\nu} \equiv \alpha \, {\cal S}^{\nu\rho\mu} + 
\beta \big( {\cal S}^{\rho\mu\nu}- {\cal S}^{\mu\rho\nu} \big) +
\gamma \big( {\cal S}_\alpha {}^{\alpha\rho} \, \eta^{\mu\nu}- {\cal S}_\alpha {}^{\alpha\mu} \, \eta^{\rho\nu} \big)
\; , \label{symmetrizing}
\ee
with arbitrary $\alpha$, $\beta$, and $\gamma$.
The only choice for which eq.~\eqref{rest} is symmetric in $\mu$ and $\nu$ is $\alpha = - \beta = 1$ and $\gamma=0$.

Thus, for this choice,  putting everything together we arrive at our Lorentz-friendly version of the stress-energy tensor:
\be
T^{\mu\nu}_{L} = T^{\mu\nu}_{c} + \frac{1}{2} \frac{\delta S}{\delta \underline{\phi}} \cdot \mathcal{J}^{\mu\nu} \cdot \underline{\phi} + \frac{1}{2} \partial_{\rho} \left[ \frac{\partial\mathcal{L} }{ \partial \, \partial_{\rho}\underline{\phi}  } \cdot \mathcal{J}^{\mu\nu} \cdot \underline{\phi} -  \frac{\partial\mathcal{L} }{ \partial \, \partial_{\mu}\underline{\phi}  } \cdot \mathcal{J}^{\rho\nu}\cdot \underline{\phi} - \frac{\partial\mathcal{L} }{ \partial \, \partial_{\nu}\underline{\phi}  } \cdot \mathcal{J}^{\rho\mu} \cdot \underline{\phi} \right] \;  . \label{Lorentz}
\ee
As we proved, as a consequence of Lorentz-invariance, this is guaranteed to be symmetric, on- and off-shell:
\be
T^{\mu\nu}_{L} = T^{\nu\mu}_{L} \qquad \qquad \mbox{(off-shell)} \;.
\ee 
It differs from the standard Belinfante expression \cite{Belinfante, Weinberg1}, which in general is symmetric only on-shell, by the second term on the r.h.s.~of \eqref{Lorentz}, which is manifestly zero on the equations of motion.

%%%%%%%%%%%%%%%%%%%%%%%%%%%%%%%%%%%%%%%
%%%%%%%%%%%%%%%%%%%%%%%%%%%%%%%%%%%%%%%

\subsection{Scale-friendly version}\label{scale friendly}
The case of scale-invariance proceeds along the same lines. Suppose that we have a scale-invariant theory in $D$ spacetime dimensions, and let's call $d$ the matrix of scaling dimensions of the fields. So, under a scale transformation $x^\mu \to (1 + \omega) \, x^\mu$, with infinitesimal, constant $\omega$, the fields transform as
\be
\underline{\phi}(x) \rightarrow \underline{\phi}(x) - \omega \, x^\mu \, \partial_{\mu} \underline{\phi}(x) - d \cdot \underline{\phi}(x),
\ee
In order to make use of this symmetry for our purposes,  we can choose $\Psi^{\mu\nu}$ in \eqref{general} to be
\be \label{Psi scale}
\underline{\Psi}^{\mu\nu}(x) = \underline{\Psi}_S^{\mu\nu}(x) = \frac{1}{D} \eta^{\mu\nu}\, d\cdot\fields
\ee
(`$S$' for `scale'),
so that we have a symmetry when $\epsilon(x )$
 takes the form appropriate for a scale transformation,
\be \label{scale epsilon}
\epsilon(x) = \epsilon_S^\mu(x) \equiv \omega \, x^\mu \; ,
\ee 
with constant $\omega$. Plugging this particular choice of $\epsilon(x )$
into \eqref{general dL}, we get
\be
\delta_S {\cal L} = - \partial_\mu (\epsilon_S^\mu {\cal L}) - \omega \, {\cal T}^{\mu} {}_\mu \; , 
\ee
while, for scale-invariance to be a symmetry, only the first term should survive. So, scale-invariance guarantees that ${\cal T}^{\mu\nu}$ is traceless, off-shell:
\be
{\cal T}^{\mu} {}_\mu = 0 \; .
\ee

Similarly to the case of Lorentz, we now have to ask whether we can choose $\Sigma^{\rho\mu\nu}$ in \eqref{general Tmn} in order to extend this property---tracelessness---to the rest of the stress-energy tensor, eq.~\eqref{rest}.
Since now ${\cal S}^{\rho\mu\nu}$ is proportional to $\eta^{\mu\nu}$ (see \eqref{Srmn} and \eqref{Psi scale}), the most 
general linear-in-${\cal S}$ combination of ${\cal S}$ and $\eta$ tensors with the right antisymmetry property is simply
\be
\Sigma^{\rho\mu\nu} = \Sigma_S^{\rho\mu\nu} \equiv \delta \big( {\cal S}^{\rho\mu\nu} - {\cal S}^{\mu\rho\nu}  \big) 
\; , \label{making traceless}
\ee
with generic $\delta$. 
Eq.~\eqref{rest} is traceless only for $\delta = -D/(D-1)$. With this choice, putting everything together we arrive at our scale-friendly stress-energy tensor:
\be \label{traceless_tensor}
T^{\mu\nu}_{S} = T^{\mu\nu}_{c} + \frac{1}{D} \, \eta^{\mu\nu} \, \frac{\delta S}{\delta \underline{\phi}} \cdot  d \cdot\underline{\phi}+\frac{1}{D-1} \partial_{\rho} \left( \eta^{\mu\nu} \frac{\partial\mathcal{L} }{ \partial \, \partial_{\rho}\underline{\phi}  }\cdot d\cdot\underline{\phi}  - \eta^{\rho\nu} \frac{\partial\mathcal{L} }{ \partial \, \partial_{\mu}\underline{\phi}  }\cdot d\cdot\underline{\phi} \right)  \; .
\ee
As a consequence of scale invariance, this is guaranteed to be traceless, on- and off-shell,
\be
T_{S}^{\mu} {}_{\mu} = 0 \qquad \qquad \mbox{(off-shell)} \;.
\ee

Before proceeding, there is a small puzzle that we need to address. According to standard results \cite{Wess, CCJ, CJ}, scale-invariance is not enough to make the stress-energy tensor traceless. The best one can do, usually, is to make it traceless, on-shell, up to a total divergence:
\be \label{standard result}
T^\mu {}_\mu = -\partial_\mu V^\mu \qquad \qquad \mbox{(standard result)} \;. 
\ee 
Here $V^\mu$ is a quantity known as the `virial current', which we will encounter and explore further in the next section. Only if the theory  enjoys full conformal symmetry can the stress-tensor  be further improved to eliminate this total divergence. In our case, there is no sign of this: scale-invariance alone guarantees that eq.~\eqref{traceless_tensor} is completely traceless, off-shell. How is this possible?

The resolution of the puzzle is that eq.~\eqref{traceless_tensor} is not symmetric in general. The standard result \eqref{standard result} assumes that one is only considering symmetric stress-energy tensors, like the Belinfante one \cite{Belinfante, Weinberg1}, which is symmetric on-shell, or our version \eqref{Lorentz}, which is symmetric off-shell as well. What we just proved is that if one gives up this requirement, in a scale-invariant theory one can make the stress-energy tensor completely traceless, off-shell. 
In fact, with hindsight, this is obvious already from eq.~\eqref{standard result}: by adding to $T^{\mu\nu}$ the trivially conserved improvement term
\be
\frac{1}{D-1}\big( \eta^{\mu\nu} \partial_\alpha V^\alpha - \partial^\nu V^\mu \big) \; ,
\ee
one can cancel the trace of $T^{\mu\nu}$ at the expense of giving up its being symmetric.

%%%%%%%%%%%%%%%%%%%%%%%%%%%%%%%%%%%%%%%%
%%%%%%%%%%%%%%%%%%%%%%%%%%%%%%%%%%%%%%%%
\subsection{Conformal-friendly version}\label{conformal friendly}
Before moving on to the case of full conformal invariance, it is instructive to first combine the two strategies that we adopted  in the previous  subsections. This will also shed some light on the tension between tracelessness and symmetry 
of the stress-energy tensor we just alluded to.

Consider then a Lorentz-invariant, scale-invariant theory. To make use of both symmetries, we can combine the $\Psi^{\mu\nu}$'s in \eqref{Psi Lorentz} and \eqref{Psi scale}, and use for the translation Noether theorem the transformation rule \eqref{general} with
\be \label{combined}
\underline \Psi^{\mu\nu}(x) = \underline \Psi_S^{\mu\nu}(x) + \underline \Psi_L^{\mu\nu}(x) \; .
\ee
In this case, eq.~\eqref{general} reduces to a translation for constant $\epsilon$, to a Lorentz transformation for $\epsilon(x)$ as in \eqref{Lorentz epsilon}, and to a scale transformation for $\epsilon(x)$ as in \eqref{scale epsilon}.

%
%
%To specialize to the case of special conformal symmetry, $\epsilon^{\mu}(x)$ should take the specific form,
%\begin{align}
%\epsilon^{\mu}(x) &= -b^{\nu} \left( 2x^{\mu}x^{\nu} - g^{\mu\nu}x^2 \right) \label{con_no_deriv_epsilon}\\
%-\partial_{\mu}\epsilon^{\nu} &= 2\left( b_{\mu}x^{\nu} - b^{\nu}x_{\mu} \right) + 2x^{\rho}b_{\rho}\delta^{\nu}_{\mu} \label{con_one_deriv_epsilon}\\
%-\partial_{\rho}\partial_{\mu}\epsilon^{\nu} &= 2b_{\mu}\delta^{\nu}_{\rho} + 2b_{\rho}\delta^{\nu}_{\mu} - 2b^{\nu}\eta_{\rho\mu}. \label{con_two_deriv_epsilon}
%\end{align}
%The change in Lagrangian density for generic $\epsilon^{\mu}(x)$ now is
%\begin{align}
%\delta \mathcal{L}  =& -\partial_{\mu}\left( \epsilon^{\mu}\mathcal{L} \right) \nonumber\\
%&- \partial_{\mu}\epsilon_{\nu}\left[ T^{\mu\nu}_{c} + \frac{\delta S}{\delta \underline{\phi}} \cdot \left( \eta^{\mu\nu} \tilde d +\frac{1}{2} \mathcal{J}^{\mu\nu} \right) \cdot \underline{\phi} + \partial_{\rho} \left( \frac{\partial\mathcal{L} }{ \partial \, \partial_{\rho}\underline{\phi}  } \cdot \left(\eta^{\mu\nu} \tilde d+\frac{1}{2}\mathcal{J}^{\mu\nu} \right) \cdot \underline{\phi} \right) \right] \nonumber\\
%&- \partial_{\rho}\partial_{\mu}\epsilon_{\nu} \bigg[ \frac{\partial\mathcal{L} }{ \partial \, \partial_{\rho}\underline{\phi} } \cdot \left( \eta^{\mu\nu} \tilde d +\frac{1}{2} \mathcal{J}^{\mu\nu} \right) \cdot \underline{\phi} \bigg] \; . \label{dL conformal}
%\end{align}
%
Precisely because of the same reasons as in the last two subsections---Lorentz invariance and scale invariance---the ${\cal T}^{\mu\nu}$ contribution to the stress-energy tensor is both symmetric {\em and} traceless, off-shell:
\be
{\cal T}^{\mu\nu} = {\cal T}^{\nu\mu} \; , \qquad {\cal T}^{\mu}{}_{\mu} = 0 \; .
\ee
 The question is what to do with the rest, eq.~\eqref{rest}. We can first notice that, as far the $\mu$ and $\nu$ indices are concerned,  our  ${\cal S}^{\rho\mu\nu}$ has the same algebraic properties as our $\underline\Psi^{\mu\nu}$ in \eqref{combined}: it is made up of a scale part, which is pure trace, and a Lorentz one, which is antisymmetric:
\be
{\cal S}^{\rho\mu\nu} =  {\cal S}_S^{\rho\mu\nu} + {\cal S}_L^{\rho\mu\nu} \; , \qquad \qquad {\cal S}_S^{\rho\mu\nu} \propto \eta^{\mu\nu} \; , \qquad {\cal S}_L^{\rho\mu\nu} = - {\cal S}_L^{\rho\nu\mu}  \; . \label{S_both}
\ee
In turn, the most generic $\Sigma^{\rho\mu\nu}$ we can add is simply the combination of \eqref{symmetrizing} and \eqref{making traceless}:
\begin{align}
\Sigma^{\rho\mu\nu} &=  \Sigma_L^{\rho\mu\nu} + \Sigma_S^{\rho\mu\nu} \nonumber\\
&=  \alpha \, {\cal S}_L^{\nu\rho\mu} + 
\beta \big( {\cal S}_L^{\rho\mu\nu}- {\cal S}_L^{\mu\rho\nu} \big) +
\gamma \big( {\cal S}_{L\;\alpha} {}^{\alpha\rho} \, \eta^{\mu\nu}- {\cal S}_{L\;\alpha} {}^{\alpha\mu} \, \eta^{\rho\nu} \big) + \delta \big( {\cal S}_S^{\rho\mu\nu} - {\cal S}_S^{\mu\rho\nu}  \big). 
\end{align}
Symmetry of $\Delta T^{\mu\nu}$ requires $\alpha=-\beta=1,\;\gamma=\delta=0$ in the equation above, while traceless-ness requires $\alpha-\beta-\gamma(D-1)=0,\;\delta  = -D/(D-1)$. The two solutions of this linear algebra problem are inconsistent, meaning one can make $\Delta T^{\mu\nu}$ symmetric, {\em or} traceless, but not both. In either case, $\Delta T^{\mu\nu}$ contributes a total derivative to the stress-energy tensor. And so, in particular, if one decides to make  it symmetric, its trace will be a total divergence, in agreement with the standard result \eqref{standard result}.

Now, what happens in the case of full conformal invariance? It so happens that infinitesimal special conformal transformations are precisely a spacetime-modulated specific combination of Lorentz- and scale-transformations of the form \eqref{combined}, when $\epsilon(x)$ in \eqref{general} is taken to be
\be \label{conformal epsilon}
\epsilon^\mu(x) = \epsilon_C^{\mu}(x) \equiv b^\mu  \, x^2  -2 b\cdot x \,  x^{\mu} 
\ee
(`$C$' for `conformal'), with constant $b^\mu$. The derivatives of such an $\epsilon(x)$ are in fact a combination of a trace part and an antisymmetric one,
\be
\partial_{\mu}\epsilon_{C \, \nu} = 2\left( b_{\nu}x_{\mu} - b_{\mu}x_{\nu} \right) - 2b\cdot x \, \eta_{\mu\nu}  \; ,\label{con_one_deriv_epsilon}
\ee
as befits  a (spacetime-dependent) combination of Lorentz- and scale-transformations. So, if we use this $\epsilon^\mu(x)$ in \eqref{general dL}, the ${\cal T}^{\mu\nu}$ term vanishes because of Lorentz- and scale-invariance, and we are left with
\be \label{delta C L}
\delta_C {\cal L} =  -\partial_{\mu}\left( \epsilon^{\mu}\mathcal{L} \right)  + 2 b_\mu \big( {\cal S}^{\mu\alpha}{}_\alpha +2  \, {\cal S}_\alpha {}^{[\mu\alpha]}\big) \; .
\ee %switched \alpha and \mu in {\cal S}_\alpha {}^{[\alpha\mu]}
Reasoning as before, we would be tempted to say that, if conformal transformations are a symmetry, only the first term should survive. But this is where the subtlety we briefly alluded to in sect.~\ref{passive vs active} becomes relevant, and so we must finally address it.

In most common cases, even under a {\em passive} transformation, the action of a (classically) conformally invariant theory is not strictly invariant under conformal transformations, but changes by a boundary term. This is not necessarily related to the existence of Wess-Zumino terms, like for instance that studied in \cite{KS}. Rather, it usually happens because one is not really using the most symmetric version of the action. To make this very explicit, consider a free massless scalar field $\Phi(x)$ in four spacetime dimensions:
\be
S[\phi] = \int d^4 x \, \sfrac12 (\partial \Phi)^2 \; .
\ee
The passive version of a special conformal transformation is
\be
x^\mu \to x'^\mu = x^\mu + b^\mu  \, x^2  -2 b\cdot x \,  x^{\mu}  \; , \qquad \Phi(x) \to \Phi'(x') = \Phi(x) -2 (b \cdot x) \Phi(x) \; ,
\ee
and it is easy to check that the action above changes by a boundary term:
\be
d^4 x' \, \sfrac12 (\partial' \Phi')^2 = d^4 x \Big( \sfrac12 (\partial \Phi)^2 - b^\mu \partial_\mu \big(\Phi^2) \Big) \; .
\ee
However, if one instead starts from the equivalent action
\be
\tilde S[\phi] = - \int d^4 x \, \sfrac12 \Phi \Box \Phi \; ,
\ee
that boundary term is gone:
\be
d^4 x' \, \sfrac12 \Phi' \Box' \Phi' = d^4 x \, \sfrac12 \Phi \Box \Phi \; .
\ee

For simplicity, whenever possible, we tend to rewrite actions in a way that they only involve up to first derivatives of the fields,  and, in fact, we have assumed just that in all of our derivations above. So, if we insist on this assumption, in general for conformal transformations we have to allow that the action change by a total derivative {\em beyond} that of eq.~\eqref{d xi L}. This means that, for a conformally invariant theory, the last term in \eqref{delta C L} must be a total derivative:
\be \label{virial}
V^\mu \equiv {\cal S}^{\mu\alpha}{}_\alpha +2  \, {\cal S}_\alpha {}^{[\mu\alpha]} =  \partial_\alpha \sigma^{\alpha\mu} \; ,
\ee %switched \alpha and \mu in {\cal S}_\alpha {}^{[\alpha\mu]}
for some local functional $ \sigma^{\alpha\mu}$. $V^\mu$ is traditionally called the `virial current' \cite{CCJ, CJ}.

We can now go back to the form of the stress-energy tensor, eq.~\eqref{general Tmn}. We already saw that its ${\cal T}^{\mu\nu}$ part is symmetric and traceless, off-shell. As to the rest, eq.~\eqref{rest}, we saw in \eqref{S_both} that $\cal{S}^{\rho\mu\nu}$ is made up of the equivalent scale and Lorentz parts. The virial current \eqref{virial} relates these two parts. And so, for example
using \eqref{virial} we can eliminate the scale part, 
\be
{\cal S}^{\rho\mu\nu} = \frac{1}{D} \big[\eta^{\mu\nu} \, \partial_\alpha \sigma^{\alpha\rho}  + D {\cal S}_L^{\rho\mu\nu} - 2 {\cal S}_{L \, \alpha} {}^{\rho\alpha} \eta^{\mu\nu} \big]  \; ,
\ee%switched \alpha and \rho in  - 2 {\cal S}_{L \, \alpha} {}^{\alpha\rho} \eta^{\mu\nu}
and rewrite \eqref{rest} as\footnote{Notice that only the symmetric part of $\sigma^{\mu\nu}$ enters the stress-energy tensor.}
\begin{align}
 \Delta T^{\mu\nu}  = & -\partial_\rho \Big( {\cal S}_L^{\rho\mu\nu} - \frac2D {\cal S}_{L \, \alpha} {}^{\rho\alpha} \eta^{\mu\nu}  +  \Sigma^{\rho\mu\nu} \Big) \nonumber \\
 &  - \frac1D  \eta^{\mu\nu} \partial_\alpha\partial_\rho \sigma^{(\alpha\rho)}  \; .  \label{Delta Tmn}
\end{align}%switched \alpha and \rho in  - 2 {\cal S}_{L \, \alpha} {}^{\alpha\rho} \eta^{\mu\nu}
This expression, together with the tracelessness and symmetry of ${\cal T}^{\mu\nu}$, encodes conformal invariance at the level of the stress-energy tensor. There is no reference anymore to the transformation properties of the fields under scale transformations because, for conformal invariant theories, those are related to the fields' Lorentz transformation properties through the virial current \eqref{virial}.

The question now is whether one can choose an (antisymmetric in $\rho$ and $\mu$)
$\Sigma^{\rho\mu\nu}$ in such a way as to make $\Delta T^{\mu\nu}$  also traceless and symmetric.
For vanishing $\sigma^{\mu\nu}$, the answer is simply the same as in  the Lorentz-friendly case---see sect.~\ref{Lorentz friendly}:
\be
\Sigma^{\rho\mu\nu} =  \Sigma_L^{\rho\mu\nu}  \equiv    {\cal S}_L^{\nu\rho\mu} - \big({\cal S}_L^{\rho\mu\nu} - {\cal S}_L^{\mu\rho\nu} \big) \; .
\ee
The reason is that, as we know from that section, this choice makes \eqref{rest} symmetric, 
and the extra terms in \eqref{Delta Tmn} are already symmetric. Moreover, as to the trace,
we have
\be
\Sigma_L^{\rho\mu}{}_\mu  = 2  \,  {\cal S}_{L \, \mu} {}^{\rho\mu}  \; ,
\ee
which makes the trace of the first line in \eqref{Delta Tmn} vanish. 

For nonvanishing $\sigma^{\mu\nu}$, we can supplement $\Sigma_L^{\rho\mu\nu}$ with total derivative terms, which, in order not to spoil the $\mu\nu$ symmetry just obtained, and recalling that $\Sigma^{\rho\mu\nu}$ must be antisymmetric in $\rho$ and $\mu$  and that it is acted on by a $\partial_\rho$ in \eqref{Delta Tmn}, should take the form
\be
\Sigma^{\rho\mu\nu} = \Sigma_L^{\rho\mu\nu} + \partial_\alpha \Xi^{[\rho \mu][\alpha \nu]} \; ,
\ee
where $\Xi$ should be symmetric under the  $\rho\mu \leftrightarrow \alpha\nu$ pair-exchange, while we are displaying the needed antisymmetries explicitly. The trace of \eqref{Delta Tmn} then is
\be \label{final trace}
\Delta T^{\mu} {}_{\mu} = - \partial_\alpha \partial_\rho \big(\eta_{\mu\nu} \, \Xi^{[\rho \mu][\alpha \nu]} + \sigma^{(\alpha\rho)} \big) \; .
\ee
Following the same logic as before, we parametrize $\Xi$ as the most general tensor with the right symmetries and constructed out of $\sigma$ and $\eta$ tensors:
\begin{align}
\Xi^{[\rho \mu][\alpha \nu]} = & \; \; A \big(\eta^{\mu\nu} \sigma^{(\alpha\rho)} + \eta^{\alpha\rho} \sigma^{(\mu\nu)} -  \eta^{\mu\alpha} \sigma^{(\nu\rho)}  -   \eta^{\nu\rho} \sigma^{(\mu\alpha)} \big) \\
& + B \big( \eta^{\alpha\rho} \eta^{\mu\nu} -  \eta^{\mu\alpha} \eta^{\nu\rho}   \big) \sigma^{\beta} {}_{\beta}  \; .
\end{align}
Demanding that eq.~\eqref{final trace} vanish, we get $A= - \frac{1}{D-2}$ and $B = \frac{1}{(D-1)(D-2)}$.

Putting everything together, we find that the conformal-friendly version of the stress-energy tensor is:
%\begin{align}
%T^{\mu\nu}_{C} =&\;\;\;\; T^{\mu\nu}_{c} + \frac{\delta S}{\delta \underline{\phi}} \cdot \left( \frac1D \eta^{\mu\nu}d + \frac{1}{2} \mathcal{J}^{\mu\nu} \right) \cdot \underline{\phi} + \partial_{\rho} \left[ \frac{\partial\mathcal{L} }{ \partial \, \partial_{\rho} \underline{\phi}  } \cdot\left(\frac{1}{D} \eta^{\mu\nu}d + \frac{1}{2} \mathcal{J}^{\mu\nu} \right) \cdot \underline{\phi} \right] \nonumber \\
%& +\frac{1}{2} \partial_\rho\left[ \frac{2}{D} \frac{\partial\mathcal{L} }{ \partial \left( \partial_{\sigma}\underline{\phi} \right) }  \mathcal{J}^{\rho}_{\;\;\;\sigma}\underline{\phi}\eta^{\mu\nu} -  \frac{\partial\mathcal{L} }{ \partial \left( \partial_{\mu}\underline{\phi} \right) } \mathcal{J}^{\rho\nu}\underline{\phi} -   \frac{\partial\mathcal{L} }{ \partial \left( \partial_{\nu}\underline{\phi} \right) } \mathcal{J}^{\rho\mu}\underline{\phi} \right] \nonumber \\
%&- \left[ - \frac{2}{D(D-2)} \eta^{\mu\nu} \partial_{\rho}\partial_{\sigma}\sigma^{(\rho\sigma)} + \frac{1}{D-2} \left( \partial^{\nu}\partial_{\rho}\sigma^{(\rho\mu)} + \partial^{\mu}\partial_{\rho}\sigma^{(\rho\nu)} \right) - \frac{1}{D-2}\Box\sigma^{(\mu\nu)} \right. \nonumber \\
%& \left. \;\;\;\;\;- \frac{1}{(D-1)(D-2)}\partial^{\mu}\partial^{\nu}\sigma^{\rho}_{\;\;\rho} + \frac{1}{(D-1)(D-2)} \eta^{\mu\nu}\Box\sigma^{\rho}_{\;\;\rho} \right] \label{conformal} \; ,
%\end{align}
% ~~~~ when S^\rho\mu\nu is expanded as in 67 instead of using S_S + S_L ~~~~
\begin{align}
T^{\mu\nu}_{C} =&\;\;\;\; T^{\mu\nu}_{c} + \frac{\delta S}{\delta \underline{\phi}} \cdot \left( \frac1D \eta^{\mu\nu}d + \frac{1}{2} \mathcal{J}^{\mu\nu} \right) \cdot \underline{\phi} 
+\frac{1}{2} \partial_\rho\left[ \frac{\partial\mathcal{L} }{ \partial \left( \partial_{\rho}\underline{\phi} \right) } \mathcal{J}^{\mu\nu}\underline{\phi} -\frac{\partial\mathcal{L} }{ \partial \left( \partial_{\mu}\underline{\phi} \right) } \mathcal{J}^{\rho\nu}\underline{\phi} -   \frac{\partial\mathcal{L} }{ \partial \left( \partial_{\nu}\underline{\phi} \right) } \mathcal{J}^{\rho\mu}\underline{\phi} \right] \nonumber \\
&+ \left[ \frac{1}{D-2} \eta^{\mu\nu} \partial_{\rho}\partial_{\sigma}\sigma^{(\rho\sigma)} - \frac{1}{D-2} \left( \partial^{\nu}\partial_{\rho}\sigma^{(\rho\mu)} + \partial^{\mu}\partial_{\rho}\sigma^{(\rho\nu)} \right) + \frac{1}{D-2}\Box\sigma^{(\mu\nu)} \right. \nonumber \\
& \left. \;\;\;\;\;+ \frac{1}{(D-1)(D-2)}\partial^{\mu}\partial^{\nu}\sigma^{\rho}_{\;\;\rho} - \frac{1}{(D-1)(D-2)} \eta^{\mu\nu}\Box\sigma^{\rho}_{\;\;\rho} \right] \label{conformal} \; ,
\end{align}
where $\sigma^{\mu\nu}$ is related to the virial current by $\eqref{virial}$, $V^\mu = \partial_\alpha \sigma^{\alpha \mu}$. This stress-energy tensor is guaranteed to be symmetric {\em and} traceless, off-shell:
\be
T_C^{\mu\nu} = T_C^{\nu\mu} \; , \qquad T_C^{\mu} \, _\mu = 0   \qquad \qquad \mbox{(off-shell)} \;.
\ee

 Before we conclude, notice that $(D-2)$ is showing up in denominators in many terms. Indeed, our procedure does not work for $D=2$, and this special case needs to be dealt with separately. It is also important that the virial current be a total derivative, as per eq.~(\ref{virial}). The reason is that in \eqref{conformal} there are combinations like $\Box \sigma^{(\mu\nu)}$ and $ \sigma^\rho {}_\rho$. These cannot be written directly in terms of the virial current $V^\mu$---rather, one needs to extract the $\sigma$ tensor from $V^\mu= \partial_\alpha \sigma^{\alpha\mu}$.

%%%%%%%%%%%%%%%%%%%%%%%%%%%%%%%%%%%%%%%%%%%
%%%%%%%%%%%%%%%%%%%%%%%%%%%%%%%%%%%%%%%%%%%

\section{The other currents, for free}

We have seen how modifying the translation Noether theorem along the lines of sect.~\eqref{generalities} can make the derivation of ``improved" stress-energy tensors systematic. The reason the strategy works is that it makes explicit use of the fact that Lorentz, scale, and conformal transformations can be thought of as suitably modulated translations. They are still defined by certain {\em constant} parameters, respectively $\omega_{\mu\nu}$, $\omega$, and $b_\mu$. 
Now, suppose we wanted to run the Noether's theorem not for translations, but for those extra spacetime symmetries. These parameters would have to be modulated in $x$ in an arbitrary way to find the corresponding currents.
But the advantage of how we organized our Lorentz-friendly, scale-friendly, and conformal-friendly {\em translation} Noether's theorems is that these can be used directly also as Noether's theorems for, respectively, Lorentz, scale, and conformal transformations
\footnote{Similar ideas were exploited in \cite{Galilean} in the case of Galilean invariance.}.

Concretely, adopting the general transformation law \eqref{general} for completely generic $\epsilon^\mu(x)$, the Lagrangian changes as in \eqref{define Tmn}. This can be taken as the definition of the stress-energy tensor associated with this particular implementation of the Noether's theorem. If we are running the Lorentz-friendly version of the theorem (sect.~\ref{Lorentz friendly}), and we perform an $x$-modulated {\em Lorentz} transformation, 
\be
\epsilon^\mu(x) = \omega^{\mu} {}_\nu (x) x^\nu \; , \qquad \omega_{\mu\nu} (x)= -\omega_{\nu\mu} (x) \; ,
\ee 
the variation of the Lagrangian density reduces to
\begin{align}
\delta {\cal L} & = - \partial_{\mu}\big(   \omega_{\nu \alpha} x^\alpha \big) T^{\mu\nu}_L + \mbox{total derivatives} \\
&  = - \partial_{\mu}\omega_{\nu \alpha} \, x^\alpha T^{\mu\nu}_L + \mbox{total derivatives}
\; ,
\end{align}
where we used that, as a consequence of Lorentz invariance, $T^{\mu\nu}_L$ is symmetric, off-shell.
Now, by definition, whatever multiplies  the derivatives of the $\omega_{\mu\nu}$ parameters in $\delta {\cal L}$ is the Noether current $M^{\mu\nu\alpha}$ associated with {\em Lorentz} transformations,
\be 
\delta {\cal L}  = - \partial_{\mu}\omega_{\nu \alpha} \, M^{\mu\nu\alpha} + \mbox{total derivatives} \; .
\ee
Taking into account that the $\omega$'s are antisymmetric, we thus have
\be
M^{\mu\nu\alpha} = \sfrac12 \big( x^\alpha T^{\mu\nu}_L -x^\nu T^{\mu\alpha}_L  \big) 
\; , \qquad \qquad \partial_\mu M^{\mu\nu\alpha} = 0 \quad \mbox{(on-shell)} \;,
\ee
in agreement with the standard result \cite{Belinfante, Weinberg1}.

Likewise, if we are running the  scale-friendly version of the theorem (sect.~\eqref{scale friendly}), and we perform an $x$-modulated {\em scale} transformation, 
\be
\epsilon^\mu(x) = \omega (x) x^\nu \; ,
\ee
the variation of the Lagrangian reduces to
\begin{align}
\delta {\cal L} & = - \partial_{\mu}\big(   \omega \, x_\nu \big) \,  T^{\mu\nu}_S + \mbox{total derivatives} \\
&  = - \partial_{\mu}\omega \, x_\nu T^{\mu\nu}_S + \mbox{total derivatives}
\; ,
\end{align}
where we used that, as a consequence of scale invariance, $T^{\mu\nu}_S$ is traceless, off-shell.
By definition, whatever multiplies the derivatives of $\omega(x)$ is the Noether current $S^\mu$ associated with scale transformations,
\be
\delta {\cal L}  = - \partial_{\mu}\omega \, S^{\mu} + \mbox{total derivatives} \; .
\ee
We thus get
\be \label{Sm}
S^\mu = x_\nu T^{\mu\nu}_S \; , \qquad \qquad \partial_\mu S^{\mu} = 0 \quad \mbox{(on-shell)}\;. 
\ee
Related to our comments at the end of sect.~\ref{scale friendly}, notice that this differs from the standard expression of the scale current in a non-conformal theory,
\be \label{standard Sm}
S^\mu = x_\nu T^{\mu\nu} + V^\nu \qquad \qquad \mbox{(standard result)}\;.
\ee
The $V^\mu$ appearing here is precisely the same as in eq.~\eqref{standard result}. In fact, using the conservation of $S^\mu$ and of $T^{\mu\nu}$, from \eqref{standard Sm} one derives \eqref{standard result}, which shows that eq.~\eqref{standard result} is only valid on-shell.
Again, the difference between our result \eqref{Sm} and the standard one \eqref{standard Sm} stems from our using a traceless but, in general, non-symmetric stress-energy tensor.

Finally, consider the conformal-friendly case (sect.~\ref{conformal friendly}). If we choose $\epsilon^\mu(x)$ to be an $x$-modulated special conformal transformation,
\be
\epsilon^\mu(x) =  b^\mu(x)  \, x^2  -2 b(x) \cdot x \,  x^{\mu}  \; ,
\ee
the Lagrangian changes by
\begin{align}
\delta {\cal L} & = - \partial_{\mu}\big(  b_\nu  \, x^2  -2 b \cdot x \,  x_{\nu}  \big) T^{\mu\nu}_C + \mbox{total derivatives} \\
&  = - \partial_{\mu}b_{\nu} \, \big(x^2 \delta^\nu_\alpha- 2 \, x^\nu   x_{\alpha}  \big)T^{\mu\alpha}_C + \mbox{total derivatives}
\; ,
\end{align}
where we used that $T^{\mu\nu}_C$ is symmetric and traceless, off-shell. Following the same logic as above, we see that the Noether current associated with special conformal transformations is
\be
K^{\mu\nu} = \left( x^2 \delta^\nu_{\alpha}- 2 \, x^{\nu}x_{\alpha} \right)T^{\mu\alpha}_{C} \; , \qquad \qquad \partial_\mu K^{\mu\nu} = 0 \quad \mbox{(on-shell)}\;,
\ee
in agreement with the standard results \cite{Wess, CCJ, CJ}.

\section{Examples}

Let us work through a few explicit examples of the different improvements we have presented for the translation Noether theorem:
\begin{itemize} 

\item \textbf{Symmetric $T^{\mu\nu}$ from Lorentz invariance}

We can start by taking a look at the non-trivial case of a Dirac spinor $\psi(x)$, with  Lagrangian
\be
\mathcal{L}=\bar{\psi}\big( i \slashed{\partial}-m\big)\psi \; .
\ee
The canonical stress-energy tensor is as usual not symmetric, 
\be
T^{\mu\nu}_c = i\bar\psi\gamma^\mu\partial^\nu\psi - \eta^{\mu\nu}\mathcal{L} \; .
\ee

Applying our formula \eqref{Lorentz} from the Lorentz-friendly procedure, we get:
\be
T^{\mu\nu}_L = i\bar\psi\gamma^\mu\partial^\nu\psi-\eta^{\mu\nu}\mathcal{L} - \frac{i}{2}\bar\psi\left[\mathcal{J}^{\mu\nu} , \gamma^\rho\right] \partial_\rho \psi -  \frac{i}{2}\partial_\rho\left(\bar\psi\gamma^\mu \mathcal{J}^{\rho\nu}\psi + \bar\psi\gamma^\nu \mathcal{J}^{\rho\mu}\psi \right) \; ,
\ee
with the Lorentz generators given by
\be 
\mathcal{J}^{\mu\nu}=\frac{1}{4}\left[\gamma^\mu , \gamma^\nu\right] \; .
\ee

The commutator $\left[\mathcal{J}^{\mu\nu} , \gamma^\rho \right] = \left(\gamma^\mu \eta^{\nu\rho} - \gamma^\nu\eta^{\rho\mu}\right)$ fixes the non-symmetric part coming from the first term in $T^{\mu\nu}_L$, while the rest of the expression is already symmetric. With a bit of $\gamma$-matrix algebra, the final expression becomes

\be
T^{\mu\nu}_L = \frac{i}{2} \bar\psi\, \gamma^{(\mu}\partial^{\nu)}\psi -\frac{i}{2} \partial^{(\mu}\bar\psi\, \gamma^{\nu)}\psi   
+\frac{i}{2}\eta^{\mu\nu}\partial_\rho\left(\bar\psi\gamma^\rho\psi\right) -\eta^{\mu\nu}\mathcal{L},
\ee

%\be
%T^{\mu\nu}_L = i \, \bar\psi\, \gamma^{(\mu}\partial^{\nu)}\psi  -\eta^{\mu\nu}\mathcal{L}
%-\frac{i}{2}\partial^{(\nu}\big(\bar\psi\, \gamma^{\mu)}\psi\big)+\frac{i}{2}\eta^{\mu\nu}\partial_\rho\left(\bar\psi\gamma^\rho\psi\right),
%\ee

which is manifestly symmetric, off-shell, as promised. 

\item \textbf{Traceless $T^{\mu\nu}$ from scale invariance}

For our scale-invariance example we may look at scalar theories of the form
\be  \label{f}
\mathcal{L} = \phi^4 f\left(\frac{(\partial\phi)^2}{\phi^4}\right),
\ee
which are scale-invariant for any function $f$ in $D=4$. The canonical stress-energy tensor for such theories is
\be 
T^{\mu\nu}_c = 2f'\partial^\mu\phi \, \partial^\nu\phi-\eta^{\mu\nu}f\phi^4,
\ee
where the prime means $f' = \partial_X f, X \equiv \frac{(\partial\phi)^2}{\phi^4}$. Notice that $T_c^{\mu\nu}$ happens to be symmetric, since we are dealing with a scalar field, but is not traceless in general. 

Our prescription for a traceless stress-energy tensor \eqref{traceless_tensor} (with $d \to 1$) gives the following improved expression for such theories:
\be 
T^{\mu\nu}_S = \sfrac{4}{3}\, f' \, \partial^\mu\phi \, \partial^\nu\phi -\sfrac{1}{3} \eta^{\mu\nu}f' \, (\partial\phi)^2+\sfrac{1}{6}\eta^{\mu\nu}\, \phi \, \partial_\rho\left(f'\partial^\rho\phi\right)-\sfrac{2}{3}\,\phi \, \partial^\nu\left(f'\partial^\mu\phi\right) \; ,  \label{example TSmn}
\ee
for which one can readily check that the trace vanishes off-shell. 

However, $T^{\mu\nu}_S $ is not symmetric in general: the first three terms are manifestly symmetric, but the last one is  symmetric only if $f'$ is a constant, that is, only if 
\be
f(X) = {\rm const} + {\rm const} \times X  \; .
\ee
It is clear from  \eqref{f} that this choice corresponds, in $D=4$, to a conformally invariant theory.

\item \textbf{Traceless, symmetric $T^{\mu\nu}$ from conformal invariance} 

Up to changing the normalization of $\phi$, the conformally invariant theory mentioned above is
\be \label{L conformal}
\mathcal{L} = \sfrac{1}{2}\left(\partial\phi\right)^2 -\lambda\phi^4 \; .
\ee
For such a simple theory, the virial term \eqref{virial} takes the form $V^\mu = {\phi} \, \partial^\mu{\phi}$ and hence $\sigma^{\mu\nu}=\frac{1}{2}\eta^{\mu\nu}{\phi}^2$.
Following \eqref{conformal}, and given that $\mathcal{J}^{\mu\nu} = 0$ for scalar fields, the improved stress-energy tensor in $D=4$ becomes
\be \label{our TC}
T^{\mu\nu}_C = \sfrac{2}{3}\, \partial^\mu{\phi} \, \partial^\nu{\phi} -\sfrac{1}{6}\eta^{\mu\nu} \left(\partial{\phi}\right)^2 +\sfrac{1}{12}\eta^{\mu\nu}{\phi}\, \Box{\phi} -\sfrac{1}{3}{\phi} \, \partial^\mu\partial^\nu{\phi} \; ,
\ee
which is manifestly symmetric and traceless, off-shell. In fact, recalling that compared to our previous example now we have $f' = \sfrac12$, we see our scale-friendly $T^{\mu\nu}_S$ in \eqref{example TSmn} reduces precisely to our conformal-friendly $T^{\mu\nu}_C$ above.

Notice that, compared to the more standard improved stress-energy tensor associated with \eqref{L conformal},
\begin{align}
T^{\mu\nu} & = \partial^\mu{\phi} \, \partial^\nu{\phi} - \sfrac12 \eta^{\mu\nu}(\partial \phi)^2 
+ \lambda \eta^{\mu\nu} \phi^4 - \sfrac{1}{6}(\partial^\mu\partial^\nu - \eta^{\mu\nu} \Box)\phi^2 
\qquad \qquad \mbox{(standard result)}\nonumber \\
& = \sfrac{2}{3}\, \partial^\mu{\phi} \, \partial^\nu{\phi} -\sfrac{1}{6}\eta^{\mu\nu} \left(\partial{\phi}\right)^2 + \lambda \eta^{\mu\nu} \phi^4 + \sfrac{1}{3}\eta^{\mu\nu}{\phi}\, \Box{\phi} -\sfrac{1}{3}{\phi} \,  \label{standard TC} \partial^\mu\partial^\nu{\phi} \; ,
\end{align}
ours has a different coefficient for the $\phi \, \Box \phi$ term and, perhaps more surprisingly, has no sign of the potential. In particular, our $T^{\mu\nu}_C$ does not depend on $\lambda$. The reason is that, as we tried to emphasize, our expressions for improved stress-energy tensors differ from the more standard ones by terms proportional to the equations of motion. For solutions of the equations of motion, the value of potential can be related to that of  $\phi \, \Box \phi$:
\be
\lambda \phi^4 =\phi \cdot \lambda \phi^3  = - \sfrac 14 \phi\,  \Box \phi \qquad \qquad \mbox{(on-shell).} 
\ee
Using this on-shell relationship, the two expressions \eqref{our TC} and \eqref{standard TC} coincide.

%
%
%
%
%The tensor is manifestly symmetric and its trace is:
%
%\be 
%T^{\mu}_{\;\mu\; C} = \frac{2(D-2)-D(D-3)}{2(D-1)}\left(\partial{\phi}\right)^2 + (D-4)\lambda{\phi}^4,
%\ee
%
%which vanishes in $D=4$, for which ${\phi}^4$ theory is conformally invariant.
%

\item \textbf{Electromagnetism}

Finally, we may also look at the electromagnetic field's Lagrangian, which is Lorentz-, scale-, and, in $D=4$, conformal-invariant. We can then apply and compare all of the three different prescriptions. Consider then
\be 
\mathcal{L} = -\frac{1}{4}F_{\mu\nu}F^{\mu\nu} \; .
\ee
The canonical stress-energy tensor is
\be 
T_c^{\mu\nu} = \partial^\nu A_\lambda F^{\lambda\mu} +\frac{\eta^{\mu\nu}}{4}F^2 \; ,
\ee
which is neither symmetric nor traceless.

The Lorentz generators for spin-1 fields are given by $\left( \mathcal{J}^{\mu\nu}\right)^\rho {}_\sigma = \eta^{\nu\rho}\delta^\mu_\sigma- \eta^{\mu\rho}\delta^\nu_\sigma$. Our formula for Lorentz-invariance yields the manifestly symmetric stress-energy tensor
\be 
T^{\mu\nu}_L = \frac{\eta^{\mu\nu}}{4}F^2 + A_\lambda \,\partial^{(\mu} F^{\nu)\lambda} +\partial_\lambda \big( A^{(\mu} F^{\nu)\lambda} \big) \; .
\ee
%
%Notice however that the tensor is now not
%gauge-invariant\footnote{cite other paper?}.
Our scale-friendly prescription \eqref{traceless_tensor} (with $d \to \frac{D-2}{2}$) instead gives
\be
T^{\mu\nu}_S = \partial^\nu A_\lambda F^{\lambda\mu} + \frac{\eta^{\mu\nu}}{4}F^2  - \frac{D-2}{2(D-1)}\partial^\nu\left(A_\lambda F^{\lambda\mu}\right) + \frac{D-2}{2D(D-1)}\eta^{\mu\nu}\left(A_\lambda\partial_\rho F^{\lambda\rho} + D\partial_\rho A_\lambda F^{\lambda\rho} \right).
\ee
Using $\eta^\mu {}_\mu=D$ and $\partial_\mu A_\lambda F^{\lambda\mu}=-F^2/2$, we see that 
the trace vanishes, in generic $D$,
\be
T_S^\mu {}_{\mu} = 0 \; ,
\ee
without the use of equations of motion. However, notice  that this stress-energy tensor is no longer symmetric. 

As well known, in $D=4$ the theory is also conformally invariant. In fact, the virial term vanishes, and the resulting conformal-friendly stress-energy tensor is
\be
T^{\mu\nu}_C = \partial_\rho A^\mu \, \partial^\rho A^\nu - \partial^\mu A^\rho \, \partial^\nu A_\rho - A^{(\mu} \partial_\rho F^{\nu)\rho}  + \frac{\eta^{\mu\nu}}{4} A_\lambda\partial_\rho F^{\lambda\rho} \; ,
\ee
which is manifestly both symmetric and traceless, without using the equations of motion. 

Notice that {\em none} of  these stress-energy tensors for the electromagnetic field is gauge invariant. This is a common issue, and it is usually fixed, on-shell, by adding ad hoc further improvement terms. For a more constructive approach, somewhat similar to ours, see instead \cite{BLS} and references therein.

\end{itemize}

%%%%%%%%%%%%%%%%%%%%%%%%%%%%%%%%%%%%%%%%%%%
%%%%%%%%%%%%%%%%%%%%%%%%%%%%%%%%%%%%%%%%%%%
\section{Summary and concluding remarks}

Despite the dryness and length of our algebra, our strategy and findings are easy to summarize:
\begin{itemize}
\item There are ambiguities in the standard formulation of Noether's theorem. One of these is related to a modification of how the fields are {\em chosen} to transform in the case of a spacetime-modulated symmetry transformation. Since spacetime symmetries beyond translations can be thought of as suitably modulated translations, such an ambiguity can be used to one's advantage, constructively, to derive directly from the translation Noether theorem the algebraic properties of $T^{\mu\nu}$ associated with said additional spacetime symmetries.

\item We formulated this strategy in general, and applied it to the cases of Lorentz invariance, scale invariance, and conformal invariance. We reobtained the standard results, albeit with some modifications: first, our stress-energy tensors have the standard algebraic properties (symmetry and/or tracelessness) {\em off-shell}; second in the case of combined Lorentz and scale invariance, we noted a tension between tracelessness and symmetry of the stress-energy tensor. The standard choice corresponds to making the stress-energy tensor symmetric. But we showed that there is an equally valid choice in which the stress-energy tensor is traceless, off-shell, but in general non-symmetric. 

\item Since the additional spacetime symmetries are incorporated into the structure of the translation Noether theorem, this serves as Noether's theorem for those additional symmetries as well, yielding directly their associated currents in terms of the stress-energy tensor.

\end{itemize}

Our unified framework shows that the standard improvement terms that make the stress-energy tensor symmetric in the case of Lorentz invariance, and traceless in the case of scale and conformal invariance have the same origin: they are a direct consequence of the fact that all those additional spacetime symmetries are suitably modulated translations.

We have already mentioned two possible extensions of our analysis at the beginning of sect.~\ref{generalities}: the case of non-linearly realized spacetime symmetries, and the case of Lagrangians with higher-than-first derivatives of the  fields.  Another possible extension would be to push the starting point of our improved Noether's theorem, eq.~\eqref{general}, to higher orders in derivatives of $\epsilon^a(x)$. We see no obvious use for this at the moment, but maybe there is one, perhaps related to the question of non-linear realizations alluded to above.
Finally, we wonder whether the viewpoint we have put forward here can prove useful for the ongoing conversation on scale vs.~conformal invariance (see for instance the recent \cite{FHH} and references therein.)

%%%%%%%%%%%%%%%%%%%%%%%%%%%%%%%%%%%%%%%%%%%
%%%%%%%%%%%%%%%%%%%%%%%%%%%%%%%%%%%%%%%%%%%
\section*{Acknowledgements}
We thank Tomas Brauner, Giorgio Torrieri, and Kazuya Yonekura for bringing to our attention refs.~\cite{Brauner:2014aha, Brauner:2019lcb, Yonekura:2012uk}, which we had missed.  Some of our ideas had been already explored there.

Our work is partially supported by the US DOE (award number DE-SC011941) and by the Simons Foundation (award number 658906).

%%%%%%%%%%%%%%%%%%%%%%%%%%%%%%%%%%%%%%%%%%%
%%%%%%%%%%%%%%%%%%%%%%%%%%%%%%%%%%%%%%%%%%%
%\appendix
%\section{}
%

%%%%%%%%%%%%%%%%%%%%%%%%%%%%%%%%%%%%%%%%%%%
%%%%%%%%%%%%%%%%%%%%%%%%%%%%%%%%%%%%%%%%%%%
% BIBLIOGRAPHY

%\bibliography{library}{}
%\bibliographystyle{ieeetr}
%

\end{fmffile}

\end{document}